\documentclass[a4paper]{article}
\pdfoutput=1
\usepackage{amsmath}
\usepackage{amssymb}
\usepackage{layout}
\usepackage{ifthen}
\usepackage{url}
\usepackage[pdftex,colorlinks=true,linkcolor=blue]{hyperref}
\usepackage[retainorgcmds]{IEEEtrantools}
\author{Roger Sewell}
\usepackage{graphicx}
\usepackage{xr}
\usepackage{float}

\title{Bayesian analysis of Biomarker levels can predict time of
  recurrence of prostate cancer with strictly positive apparent
  Shannon information against an exponential attrition
  prior\footnote{Second version deposited in arxiv; v2 clarifies that
  Gleason grade and stage also considered; RFS version 1.9.1.1
.}}
\author{Roger Sewell\footnote{Trinity College, Cambridge; but at the
  time this work was done employed full time by Cambridge Consultants.}\\
\href{mailto:roger.sewell@cantab.net}{\scriptsize{roger.sewell@cantab.net}}\\
Elisabeth Crowe\footnote{Cancer Research UK Cancer Centre; but at the
time this work was done employed full time by Cambridge
Consultants.}\\
Sharokh F. Shariat\footnote{Department of Urology, Medical University
of Vienna.}}

\begin{document}
\maketitle

\begin{center}
\textbf{Abstract}
\end{center}

Shariat et al \cite{Shariat} previously investigated the possibility
of predicting from clinical data (including Gleason grade and staging)
and preoperative biomarkers which of any pair of patients would suffer
postoperative recurrence of prostate cancer (relapse) first. We wished
to establish the extent to which predictions of time of relapse taken
from such a model could be improved upon using Bayesian methodology.

The same dataset analysed by Shariat et al was reanalysed using a
Bayesian skew-Student mixture model. Predictions were made of which
of any pair of patients would relapse first, as in
\cite{Shariat}. Further, predictions were made of the time of relapse.
The benefit of using these biomarkers relative to predictions made
without biomarkers, i.e. prediction quality over that of the prior,
was measured by the apparent Shannon information\cite{ASI}, using as prior a
simple exponential attrition model of relapse time independent of
input variables.

Using half the dataset for training and the other half for testing,
predictions of relapse time from the Cox model interpreted strictly
gave $-\infty$ nepers of apparent Shannon information, because it
predicts that relapse can only occur at times when patients in the
training set relapsed. Deliberately smoothed predictions from the Cox
model gave $-0.001$ ($-0.131$ to $+0.120$) nepers, while the Bayesian
model gave $+0.109$ ($+0.021$ to $+0.192$) nepers (mean, $2.5$ to
$97.5$ centiles), being positive with posterior probability $0.993$
and beating the blurred Cox model with posterior probability
$0.927$. The performance of a version of the Cox model in which hazard
rate was instead assumed invariant for any single patient over time
lay between the two, giving $+0.046$ ($-0.073$ to $+0.160$)
nepers. These predictions from the Bayesian model thus outperform
those of the Cox model as expected, but the overall yield of
predictive information leaves plenty of scope for improvement of the
range of biomarkers in use.

The Bayesian model presented here is the first such model for prostate
cancer to consider the variation of relapse hazard with biomarker
concentrations to be smooth, as is intuitively believable. It is also
the first model to be shown to provide more apparent Shannon
information than the Cox model and indeed the first to be shown to
provide positive apparent information relative to an exponential
prior.

\tableofcontents

\section{Introduction}

Prostate cancer is the most commonly diagnosed cancer and the second
leading cause of cancer death in men in the United States. Up to 30\%
of patients experience at least biochemical recurrence following
initial therapy with curative intent. It is believed that identifying
this subset of prostate cancer patients at the time of initial surgery
would allow selection of a subset of patients who should receive
additional therapies in the months following initial prostatectomy,
thus avoiding the additional cost, inconvenience, and morbidity
associated with giving all patients such additional treatment.

Further, it may be possible to use such likelihood of early recurrence
when choosing patient groups to enter clinical trials for novel
therapies, shortening the duration of such trials by facilitating
selection of those patients most likely to experience early recurrence
on the basis of biomarker data.

Various authors have attempted to use combinations of biomarkers and
clinical data to predict which patients are at highest risk of
recurrence. In particular, Shariat et al\cite{Shariat} collected a
dataset of 423 patients and analysed it using standard Cox
proportional hazards methods\cite{Cox}. The present paper uses the
same dataset and analyses it using parametric Bayesian methods which
provide a predictive distribution on time of relapse for each patient
at the time of initial surgery. The aim was to determine what
improvement in results was available from applying Bayesian methods in
place of the widely-used Cox proportional hazards model.

We will refer to ``biochemical recurrence of prostate cancer'' simply
as ``relapse''.

\section{Methods}

The patients, biochemical methods, and dataset have been described
previously in full\cite{Shariat} and we here repeat this information
in abbreviated form, along with a description of the new methodology
in this paper.

\subsection{Patient population}

This study received institutional review board approval.

423 patients were treated with radical prostatectomy and bilateral
lymphadenectomy for clinically localised prostatic adenocarcinoma
between July 1994 and November 1997.  No patient received neoadjuvant
chemotherapy, hormone therapy, or radiotherapy, and none had secondary
cancers.

The patients were followed with digital rectal examinations and serum
prostate-specific antigen (PSA) measurements. Biochemical recurrence
(herein ``relapse'') was defined as a sustained elevation on two or
more occasions of serum total PSA $>$ 0.2 ng/ml and was backdated to the
first value $>$ 0.2 ng/ml. No patient received adjuvant therapy before
this.

\subsection{Biomarkers measured}

Biomarkers were measured on preoperative plasma samples collected at
least 4 weeks after transrectal needle biopsy of the prostate and
after overnight fasting, anticoagulated with sodium citrate and
centrifuged for 20 mins at 1500g. The supernatant diluted plasma was
decanted and frozen at -80 C in polypropylene cryopreservation vials.
Serum total PSA was measured with the Hybritech assay (Hybritech
Inc). Plasma TGF-$\beta 1$, IL-6, sIL-6R, VEGF, VCAM-1, and endoglin were
measured with commercial enzyme immunoassays from R\&D Systems. Plasma
uPA, PAI-1 and uPAR levels were measured with enzyme immunoassays from
American Diagnostica. Recentrifugation of thawed plasma at 10000g for
10 mins before assay was used to prevent contamination with markers
released from damaged platelets. All samples were run in duplicate and
the mean used for the measurement input to the prediction algorithm;
intra-assay precision coefficients of variation were $<$ 10\%. Gleason
grade sum and clinical stage were included in the data vector for each
patient. 

\subsection{Bayesian algorithm design}

The detailed design of the Bayesian algorithm is described in
\cite{ASI}. Briefly, log-transformed true values of the input data and
time of relapse were modelled as a Dirichlet-based mixture of
skew-Student distributions, the parameters of all components of which
were modelled using a Bayesian hierarchical model. The logarithms of
the actual observations were modelled as differing from their true
values by Student-distributed noise, with the time of either relapse
or relapse-free end of follow-up assumed additionally to be censored
at an independently and log-normally distributed time when follow-up
would have ceased in the absence of relapse. In the case of patients
who remained relapse-free throughout, this censoring was assumed to
occur at the time follow-up actually ceased, while for relapsing
patients it was known to occur at or after the time of relapse.

Given the training dataset, the parameters of the Bayesian model were
sampled using Markov chain Monte-Carlo techniques, and the result of
training encapsulated by those parameter samples. Given then a new
patient for whom a prediction of relapse time was required, the
predictive distribution of relapse time given the biomarker data was
calculated for each parameter sample, and the average of all such
predictions over parameter samples taken as the predicted distribution
of relapse time.

\subsection{Cox proportional hazards model}

The Cox proportional hazards model against which comparison was made
was constructed using the full list of biomarker and clinical
variables, as in the penultimate entry of Table 3 of \cite{Shariat},
fitted by maximum likelihood on the training set. The fitted basal
hazard function and effect coefficients of the input variables were
used to construct predictions of the hazard against time for each
individual and hence of the probability distribution of relapse time.

\subsection{Constant-hazard Cox model}

Because much of the poor performance of the Cox model appears to be
related to overfitting of the basal hazard function to the training
data, we also considered a time-independent version of the Cox model
in which the basal hazard function was constrained to being constant
over time. As a result, the predictions of probability density of
relapse time from this model are decaying exponentials with a variety
of time constants dependent on the input variables.

\subsection{Avoidance of bootstrapping}

As described in appendix \ref{bootstrapping}, bootstrapping is
inappropriate when used with complicated prediction algorithms. In
this context there is no rigorous justification for bootstrapping even
as the number of patients approaches infinity. On the other hand there
are clear examples where bootstrapping provides grossly optimistic
results (e.g. as much as 82\% correct classification when measured by
bootstrapping versus 50\%, i.e. no better than chance alone, when
correctly measured). There are also good theoretical reasons why use
of bootstrapping would be expected to cause erroneous operation of any
training methods which infer variance of mixture components in a
generative model for the data.

Accordingly we have reprocessed the data using the original Cox
proportional hazards model used by Shariat et al\cite{Shariat} but
without bootstrapping, for comparison with the Bayesian methods here
reported, also without bootstrapping. The only bootstrapped result we
will report is a direct comparison with the published result in
\cite{Shariat}, solely for the purpose of completeness and to
demonstrate that bootstrapping gives the same false impression of a
better result in our method as in the method in \cite{Shariat}.

\subsection{Measurement of prediction quality}
\label{predqual}

The prediction quality measure used in \cite{Shariat} was that of
concordance between actual order of relapse of any pair of patients
and the order of relapse considered most likely for that pair of
patients by the model. That paper did not publish either any measure
of how the confidence of the model on the predicted order of relapse
accorded with what actually transpired, or on the time of relapse,
although the Cox model is capable of providing both.

In the present paper the Bayesian methodology used provides
probabilities\cite{Lindley} that one of any pair of patients will die
first, the probability that a patient will undergo relapse within 100
months after surgery, and a probability distribution on the time of
relapse of any patient given that they do relapse by 100 months. For
simplicity, we have combined the last two to give a probability
distribution on time of relapse, in which the area under the curve
between two time points gives the probability of relapse occurring
during that interval. For such a curve the total area under the curve
from start to 100 months gives the probability of relapse before 100
months; that probability is usually less than one.

For each of these predicted probabilities we estimate the apparent
Shannon information content\cite{ASI,DJCMpersonal,Winkler,Shapiro} of
the predictions using only data not seen by the model during training,
using the methods described in \cite{ASI}, and compare it with the
same measure on the same data analysed using the Cox model with one
modification. The modification arises because the Cox model predicts
that all relapses will occur at times (relative to time of surgery) at
which patients in the training set relapsed (because it maximises the
probability of seeing the training set jointly over both the
underlying hazard function and the model coefficients). Consequently,
unless the inordinately unlikely event occurs that this is indeed what
happens to the unseen patients, the Cox model would score $-\infty$ on
this measure, providing too easy a straw man to knock down. We
therefore felt it reasonable to give the Cox model some latitude,
spreading each ``spike'' of relapse probability evenly over the range
of time nearer to that spike than to others.

\begin{figure}
\begin{center}

\includegraphics[scale=0.6]{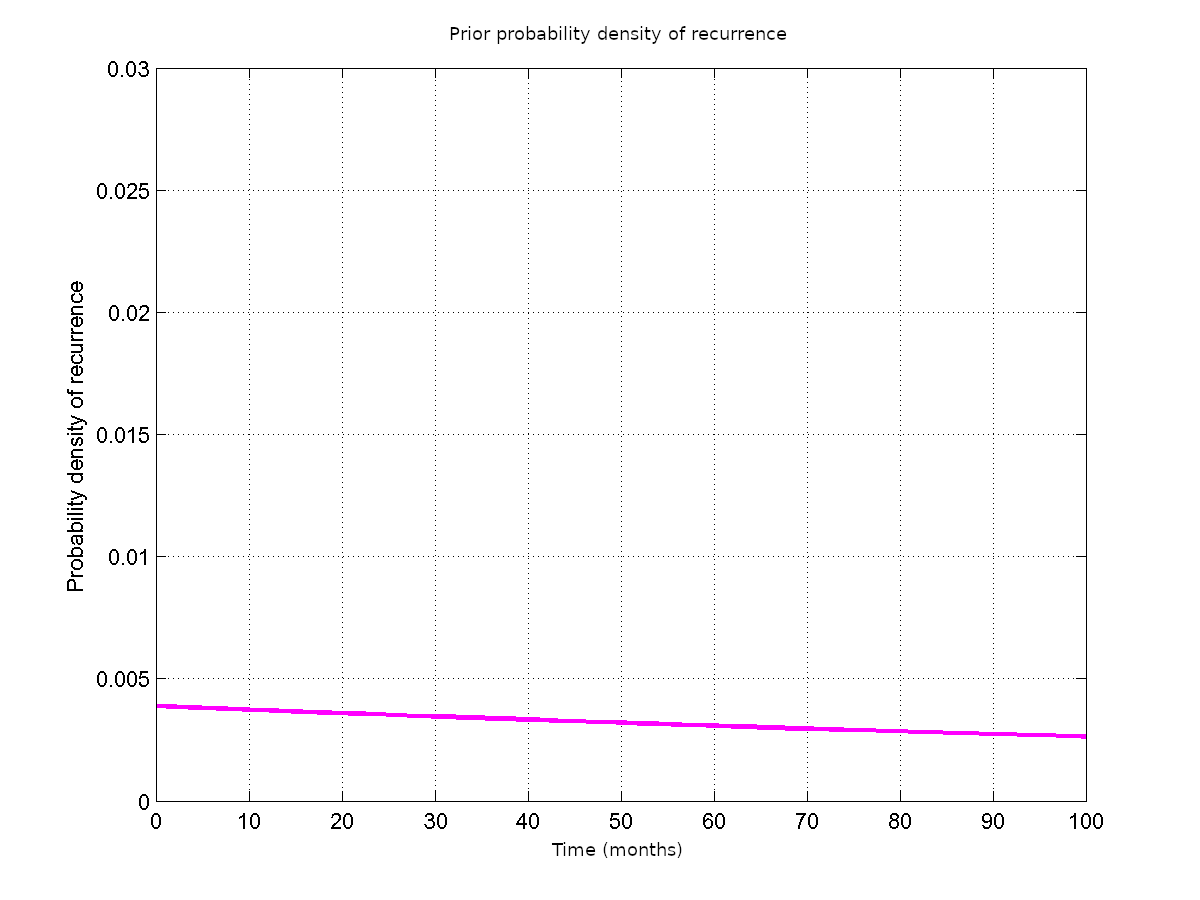}

\end{center}

\caption{Prior probability density of time of recurrence, which is an
  exponential attrition of constant hazard rate independent of
  biomarkers or clinical data.
\label{fig1}
}

\end{figure}

\begin{figure}
\begin{center}

\includegraphics[scale=0.6]{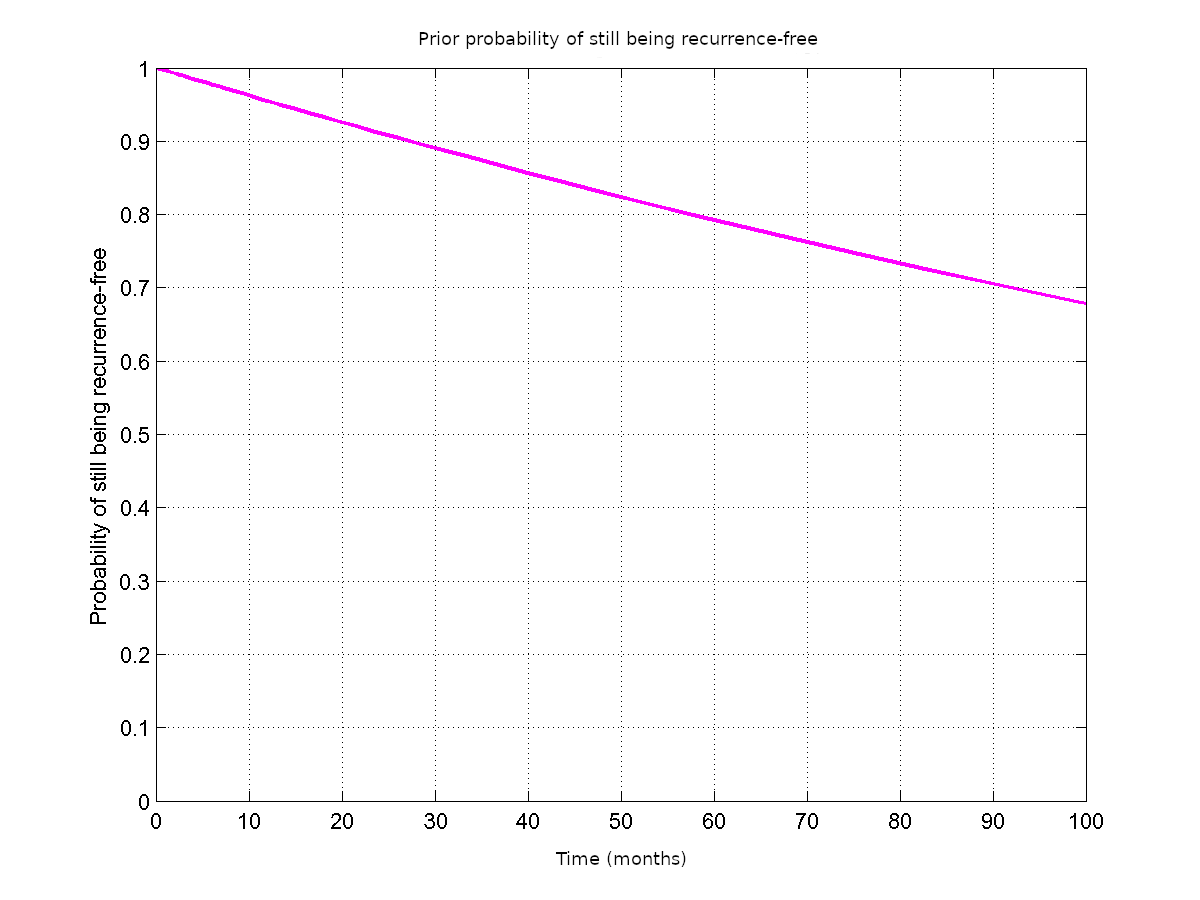}

\end{center}

\caption{The decumulative distribution corresponding to the prior probability density
shown in Figure \ref{fig1}. 
\label{fig2}
}

\end{figure}

The zero point of the Shannon information measure was set by assuming
that in the absence of input data, the hazard rate of relapse is
constant both over time and between patients. This results in an
exponential attrition model, which was fitted to the training data
answers in each case, independent of the input biomarkers and clinical
data. The resulting prior is shown in Figure \ref{fig1} as a
probability density and in Figure \ref{fig2} as the decumulative
distribution function; both will be used later for comparison in
different settings.

\subsection{Splitting the training and validation sets}
\label{splitting}

We considered three ways of splitting the training and validation
sets, namely:

\begin{enumerate}

\item Assign each patient with probability 0.5 to the validation set,
  otherwise to the training set. Train a single algorithm on the
  training set and measure its performance on the validation set. We
  refer to this as the 1/2 scenario, indicating that testing was on a
  single algorithm and that the validation set was roughly one half of
  the dataset.

\item Assign each patient with probability 1/8 to each of eight
  separate validation sets. Train eight algorithms on the eight
  different training sets, each of which consists of roughly 7/8 of
  the dataset, and which overlap with each other. Test each algorithm
  only on the points not used to train it. Assemble the resulting
  predictions into a single set of results covering all 423 patients
  exactly once. We refer to this as the 8/8 scenario.

\item Create ten overlapping validation sets, each similarly
  containing roughly $e^{-1}$ of the data. Create ten algorithms, each
  trained on one of the ten training sets containing a corresponding
  $1-e^{-1}$ of the data. Assess each algorithm only on the points not
  used to train it. Assemble the resulting predictions into a set of
  on average $10/e$ separate predictions on each datapoint. We refer
  to this as the $10/e$ scenario.

\end{enumerate}

Thus in all cases, predictions were only made on datapoints that had
not been seen during training. The 1/2 scenario provides the cleanest
assessment, and the only one performed using a single algorithm, with
the potential disadvantage that both training and validation sets are
limited to only half the available data. The 8/8 scenario achieves
training dataset size close to that of the whole dataset, with
validation on the whole dataset, but with the disadvantage that we are
no longer looking at the performance of a single algorithm. The $10/e$
scenario is at least to some extent appropriate for assessing the
variability of results with selection of training set, and also
provides input for the single measurement of bootstrapped performance
that we report.

\section{Results}

\begin{figure}
\begin{center}

\includegraphics[scale=0.6]{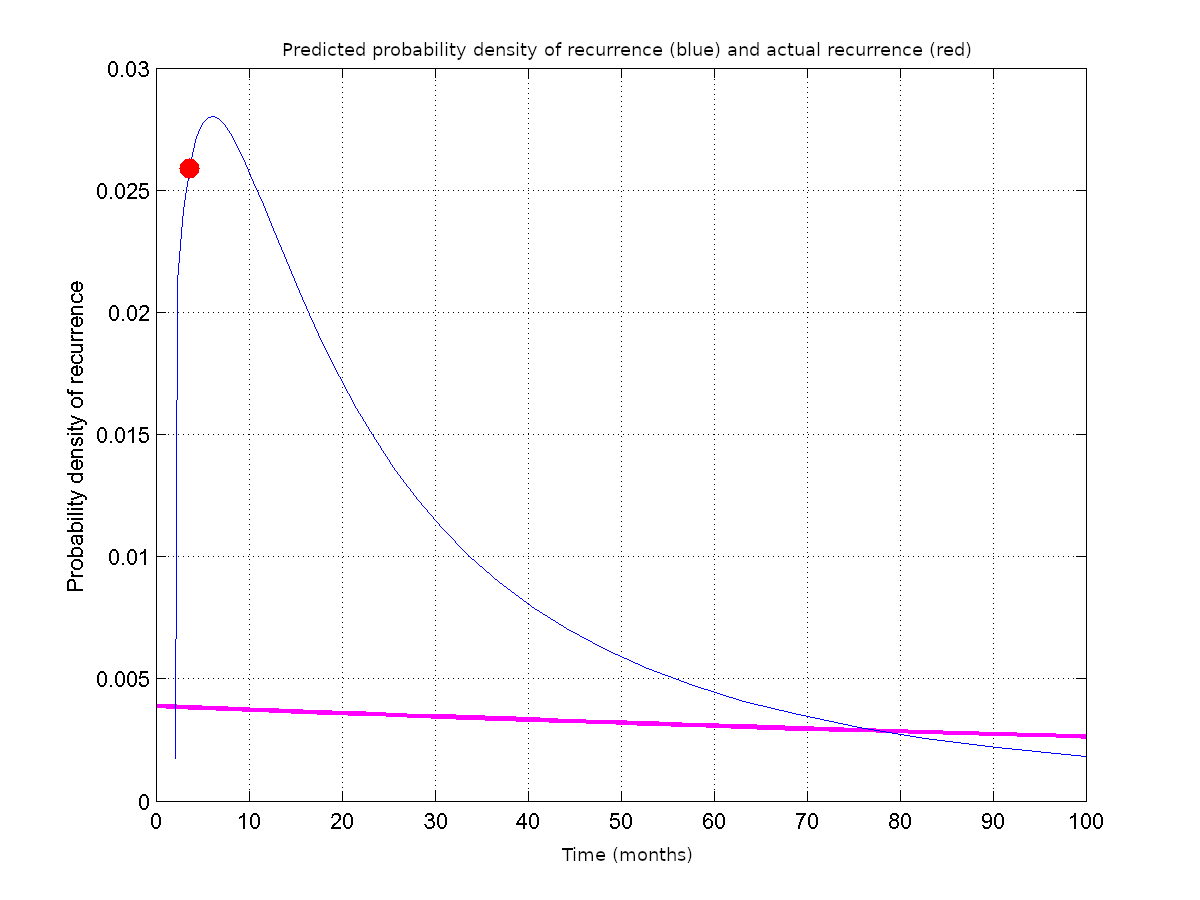}

\end{center}

\caption{The prior probability density of time of recurrence (months
  post-op) is shown in magenta (the area under which from time 0 to
  infinity is 1), with the prediction of recurrence time for a single
  patient in blue (the area under which from time 0 to infinity is
  less than or equal to 1). This particular patient did in fact
  relapse during follow up, at the time indicated by the red
  marker. The fact that the red marker is above the magenta curve
  indicates that this particular prediction carried positive apparent
  Shannon information.
\label{fig3}
}

\end{figure}

To facilitate understanding of the plots in the results, in Figure
\ref{fig3} we show the predicted probability density of time of
recurrence for one individual patient whose prediction based on
biomarkers has positive information. We also show the prior. The fact
that the prediction has value (positive apparent Shannon information)
is indicated by the fact that the red marker, indicating the actual
time of relapse, is at a point on the probability density curve that
is higher than the prior.

In the results sections that follow, we show pictures of the
predictive results made in the 1/2 test scenario, for Bayesian and Cox
models, both for patients who relapse and patient who are relapse
free. We then give in tables the results of all the other prediction
scenarios addressed.

In addition, we note that the Bayesian algorithm achieved Bootstrapped
concordance of 94\%, which compares with 86\% achieved by the Cox
proportional hazards model using the same biomarkers in
\cite{Shariat}. However, we stress that this result is not important,
and not representative of behaviour on truly unseen data, as described
in appendix \ref{bootstrapping}.

\subsection{Predictions on time of relapse}

\subsubsection{Patients observed to relapse}
\label{relapsed}

\begin{figure}
\begin{center}

\includegraphics[scale=0.6]{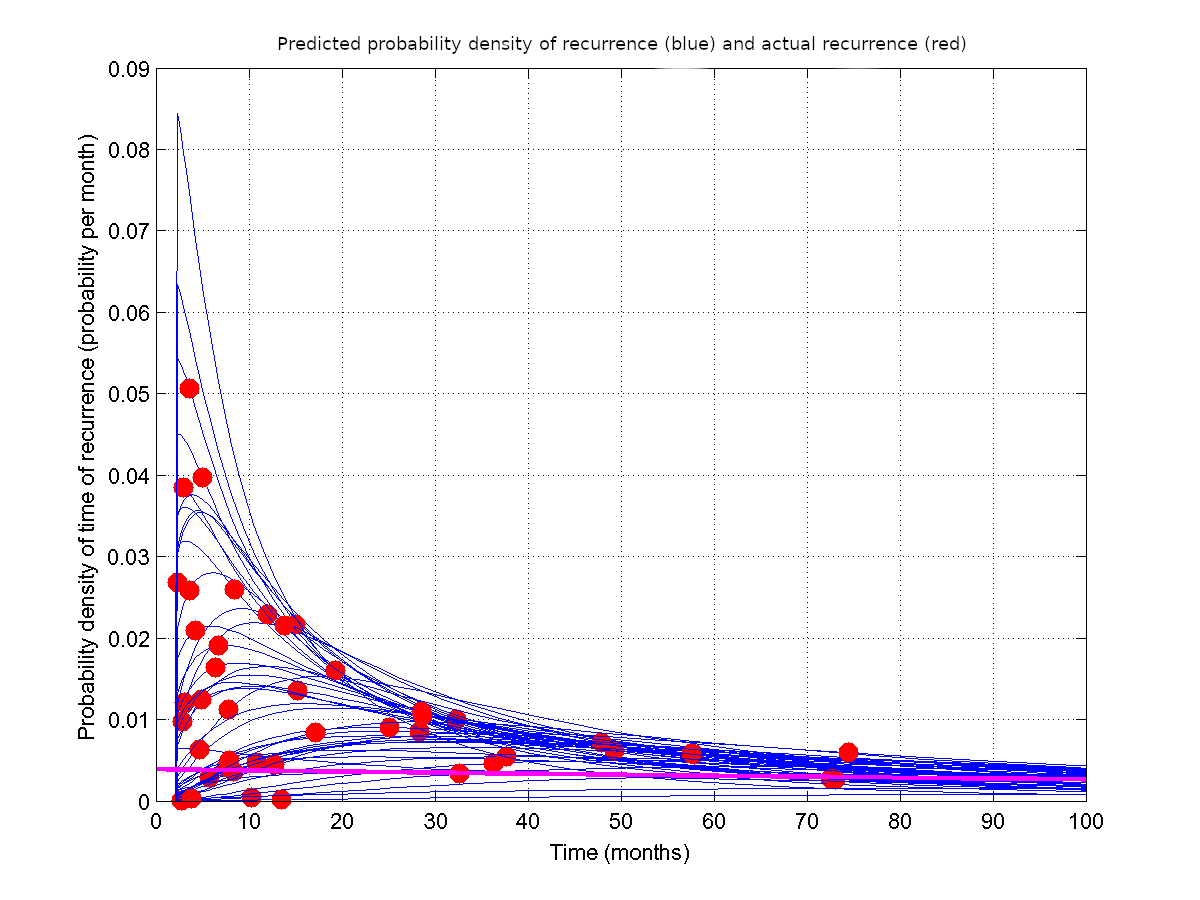}

\end{center}

\caption{All the patients in the test set, unseen during training, who
  relapsed during follow up. The magenta line is the prior probability
  density of relapse time (months post-op); the blue curves are the
  predictions of probability density of relapse time, one for each
  patient. The red blobs indicate the time when the patient actually
  relapsed.
\label{fig4}
}

\end{figure}

Figure \ref{fig4} shows all the predictions of relapse time made on
patients in the 1/2 test set who did actually relapse during
follow-up. Also shown is the prior prediction without benefit of
biomarkers or other clinical data. The blue lines indicate the
predicted probability density of relapse based on the biomarkers. The
red markers indicate the time the patient actually relapsed. Where the
prediction (red marker) has a higher probability than the prior, the
biomarker data is giving useful information; where the red marker is
below the prior, the biomarker data is misleading.

As can be seen, 36 of the 45 times the patients did actually relapse
had higher predicted probability density at that time than predicted
under the prior, showing that for these patients the predictions had
positive value (and positive apparent Shannon information). Only on
nine of the 45 patients was the biomarker-predicted density lower than
the prior density, on just four of these below three quarters of the
prior density.

\begin{figure}
\begin{center}

\includegraphics[scale=0.6]{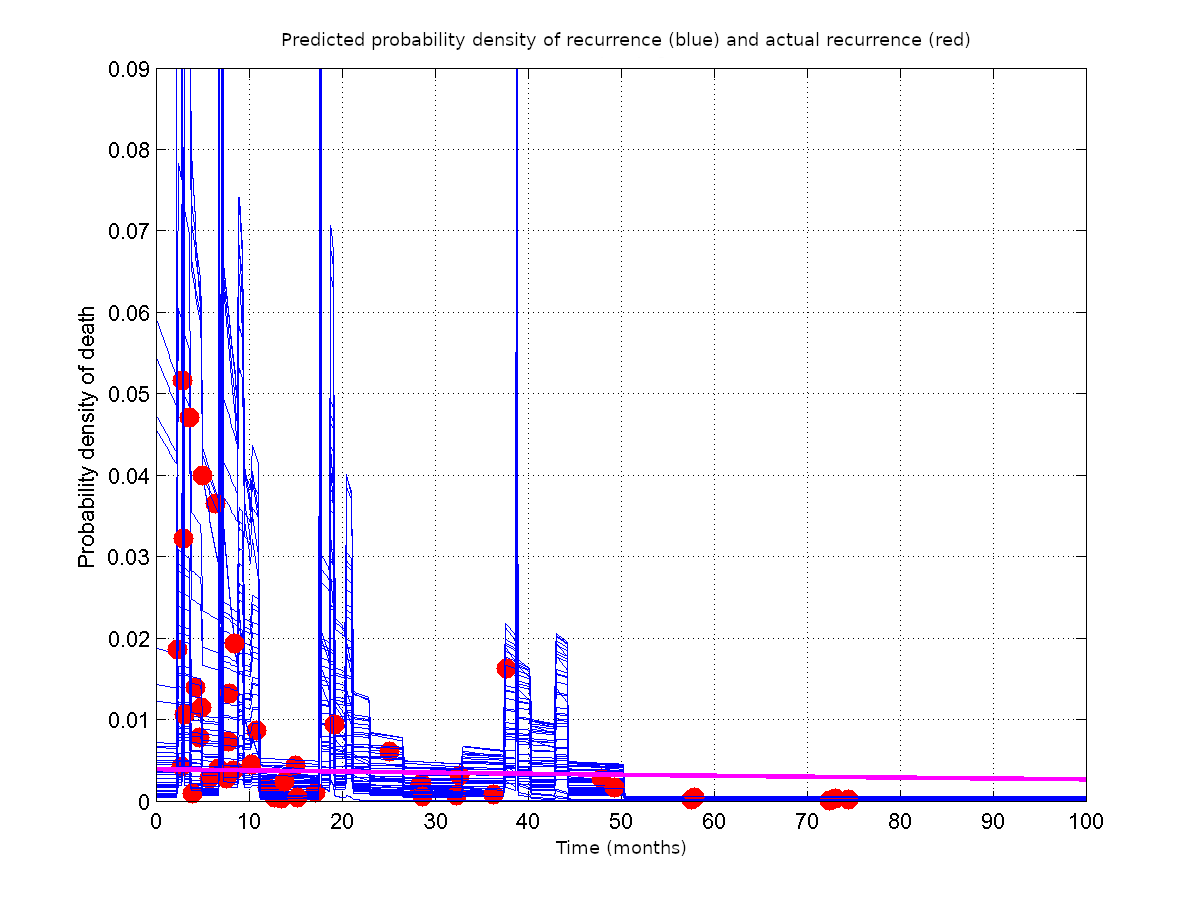}

\end{center}

\caption{Similar plot to Figure \ref{fig4} but made using the Cox
  proportional hazards model, interpreted leniently. 
\label{fig5}
}

\end{figure}

Figure \ref{fig5} in contrast shows the predictions of time of relapse
interpreted according to the Cox proportional hazards model,
interpreted leniently (spreading each ``spike'' of relapse probability
evenly over the range of time nearer to that spike than to others). A
number of features stand out. First, of the 45 patients who relapsed
during follow-up, 23 have predicted probability density at the time of
relapse lower than that of the prior, of which 20 are below
three-quarters of the prior density. Second, all patients are
predicted to have certain times which carry far higher risk of relapse
than others, for example at 17.7 months and at 38.8 months post-op.

\begin{figure}
\begin{center}

\includegraphics[scale=0.6]{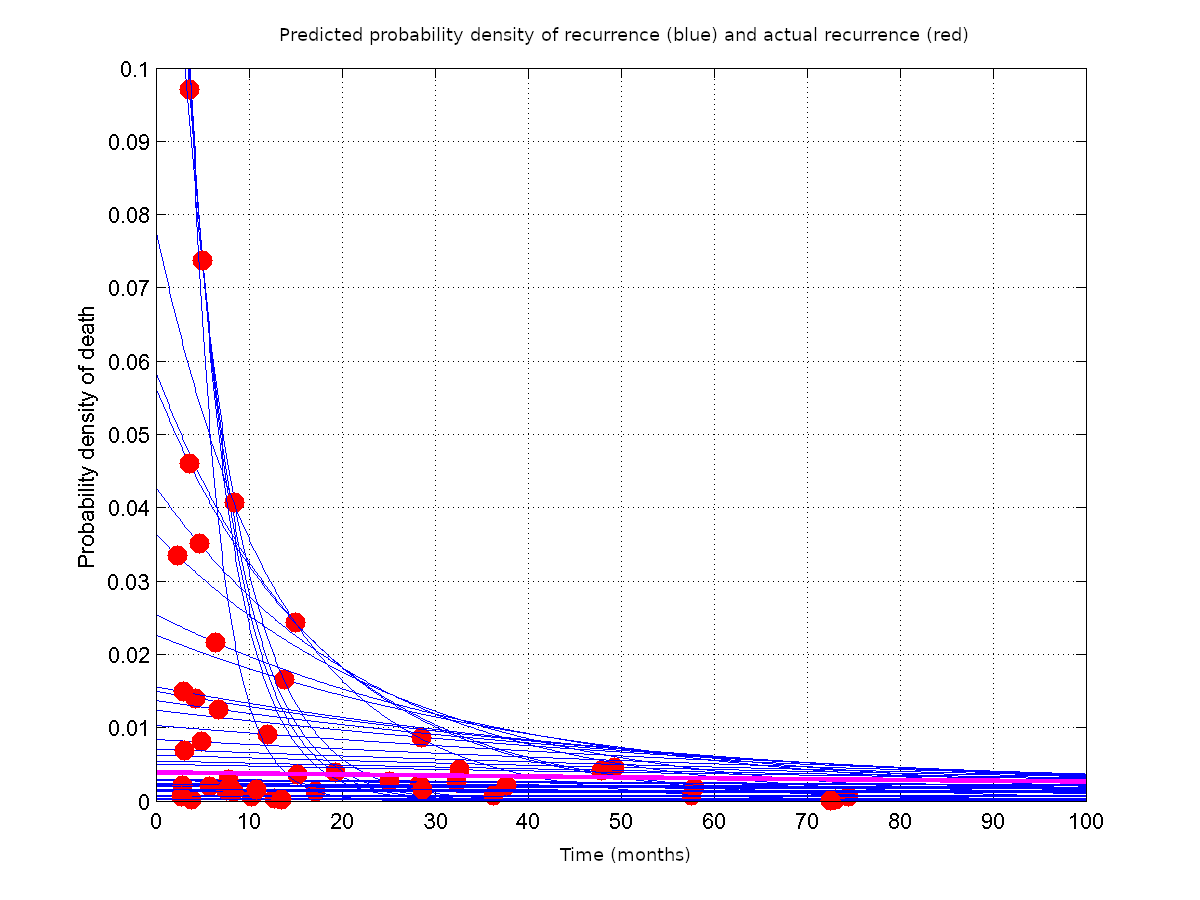}

\end{center}

\caption{Plot corresponding to Figure \ref{fig4} and Figure \ref{fig5}
  but for the constant-hazard Cox model. 
\label{fig6}
}

\end{figure}

Because some might consider that the Cox proportional hazards model's
predictions were poor because of maximum-likelihood-induced
``overtraining'', we also looked at a model similar to the Cox
proportional hazards model but in which the basal hazard rate was
assumed constant over time. The corresponding plot is shown in Figure
\ref{fig6}. 25 of the 45 patients have received worse predictions than
they could have had from the prior, 22 of them with less than
three-quarters of the density under the prior.

\subsubsection{Patients not observed to relapse}

\begin{figure}
\begin{center}

\includegraphics[scale=0.6]{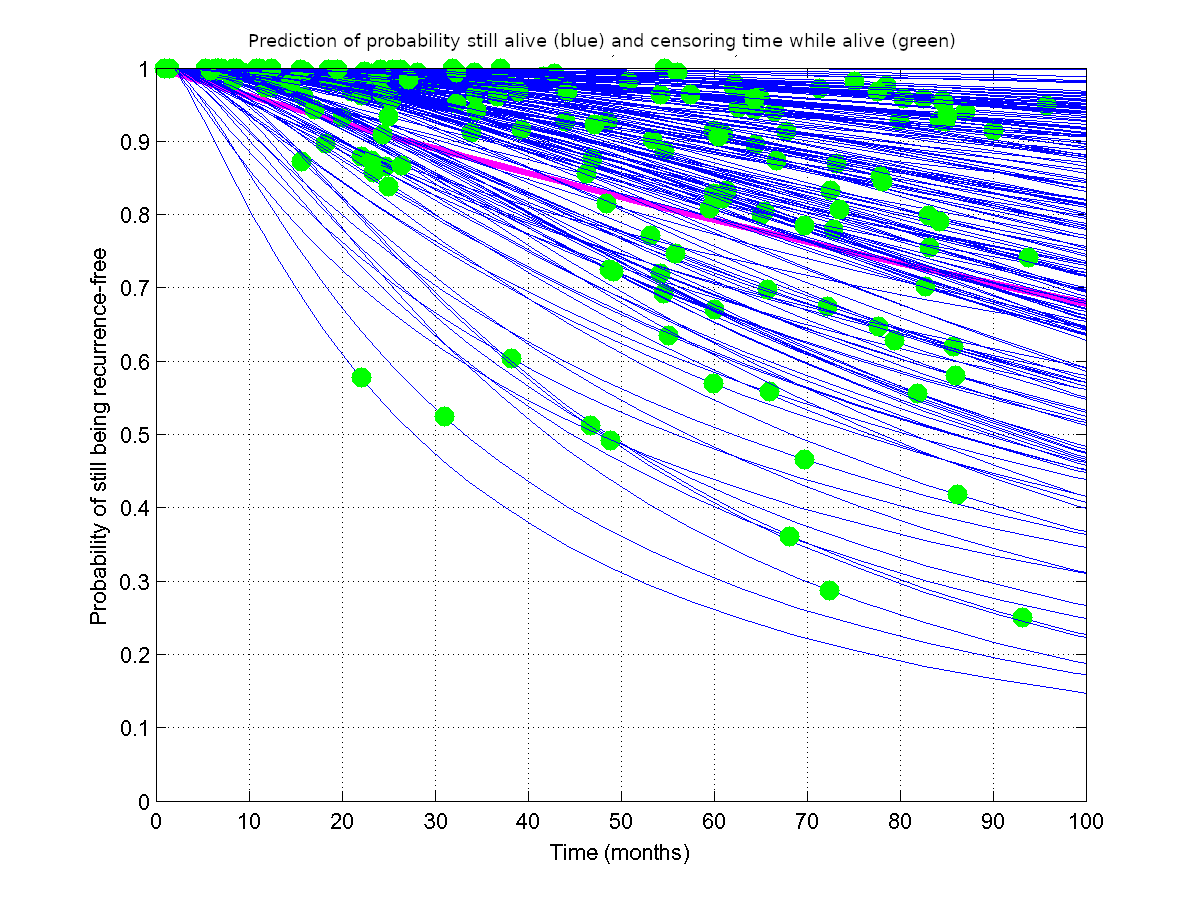}

\end{center}

\caption{Predictions of probability of still being relapse-free for
  patients who actually were relapse-free at end of follow-up. The
  magenta curve is the prior decumulative distribution, there is one
  blue curve per patient, and each green blob indicates the time at
  which follow-up for that patient ceased with the patient free of
  relapse.
\label{fig7}
}

\end{figure}

Turning attention to the remaining patients who didn't relapse during
follow-up, Figure \ref{fig7} shows the Bayesian predictions for the
remaining patients who remained relapse-free at the end of follow-up.

For 149 of the 185 patients, the predicted probability of leaving
follow-up relapse-free is greater than that under the prior, while 36
of the 185 patients have predicted probability of leaving follow-up
relapse-free below that under the prior.

\begin{figure}
\begin{center}

\includegraphics[scale=0.6]{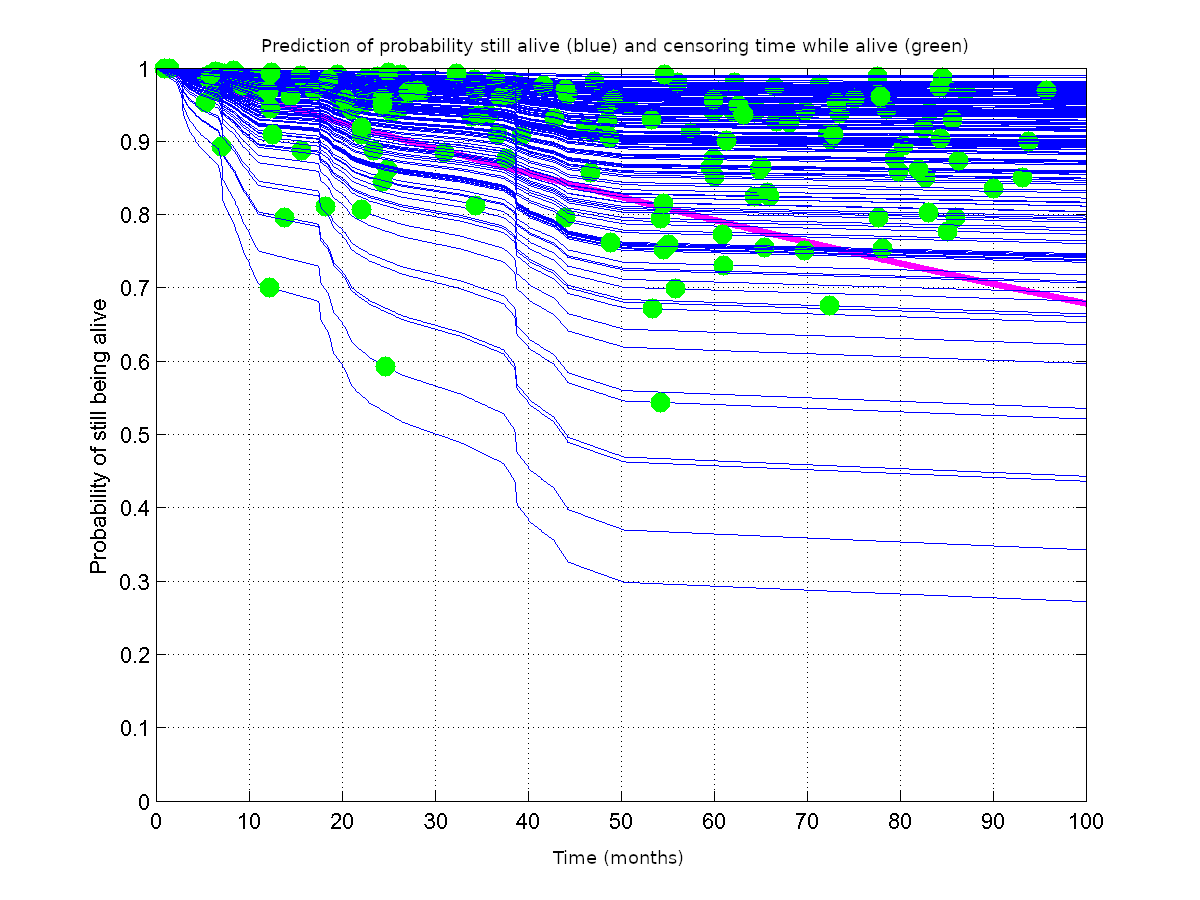}

\end{center}

\caption{Similar plot to Figure \ref{fig7} but made using predictions
  from the Cox proportional hazards model, interpreted leniently. 
\label{fig8}
}

\end{figure}

In Figure \ref{fig8} the corresponding plot for Cox model predictions,
interpreted leniently, is shown; 31 of the 185 patients have predicted
probability of leaving follow-up relapse-free below that under the
prior.

\subsubsection{Measurement of prediction quality for all patients not seen during
training}

As described in \cite{ASI}, the measure used for prediction quality is
the apparent Shannon information, which we report in nepers. Apparent
Shannon information reports the log of the geometric mean of the ratio
of the predicted probability density at time of relapse to that under
the prior (in the case of patients observed to relapse) or of the
ratio of predicted probability of still being relapse-free at the end
of follow-up to that under the prior (in the case of those not
observed to relapse). A value of $x$ nepers means that that average
ratio is $e^x$. It is positive if on average the prediction is better
than the prior.

On computing the apparent Shannon information for predictions on all
patients in the test set (not seen during training, whether or not
observed to relapse) in the 1/2 scenario, the mean posterior apparent
Shannon information content of the Bayesian predictions was +0.109
(+0.021 to +0.192) nepers (mean, 2.5 and 97.5 centiles of the
posterior distribution), positive with probability 0.993.

On the other hand the predictions of the Cox proportional hazards
contained $-0.001$ ($-0.131$ to $+0.120$) nepers of information, less
than that in the Bayesian model with posterior probability 0.927 . The
constant-hazard Cox model gave +0.046 ($-0.073$ to $+0.160$) nepers of
information, lying somewhere between the poor performance of the Cox
proportional hazards model and the better performance of the Bayesian
model.

\subsection{Predictions on order of relapse}

Simple concordance in the above scenario was 0.824 for the Bayesian
model, 0.795 for the Cox proportional hazards model, and 0.753 for the
constant-hazard Cox model.  However, concordance is calculated simply
from whether the model predicts a probability of more than 0.5 that
the patient who relapsed earlier would do so, and takes no account of
whether predictions on correct concordance were confident (near
probability 1) or unconfident (near 0.5). Apparent Shannon information
content takes appropriate account of this, and using this measure the
Bayesian model gave 0.288 (0.278 to 0.299) nepers of information about
order of relapse, the Cox model 0.181 (0.165 to 0.196), and the
constant-hazard Cox model 0.128 (0.105 to 0.151) nepers, with
posterior probabilities greater than 0.99 for all three comparisons
being in this order.

\subsection{Summary of results in other training scenarios}

Table \ref{table1} in the appendix section \ref{tables} shows the
results of analyses in the other training scenarios; caveats are
listed in the caption to the table. Comparisons between the different
methods are shown in Table \ref{table2}. The general tenor of these results is
that the Bayesian method is better than the constant hazard Cox model
which in turn is better than the standard proportional hazards Cox
model. There is, however, one possible exception: in the 8/8 scenario
only (where in each case eight different trained algorithms are being
invoked on each side of the comparison) there is a probability of
about 0.78 that the constant-hazard Cox model may give more
information than the Bayesian model about time of relapse.

\section{Discussion}

Bayesian inference provides optimal inference and prediction for any
completely mathematically specified inference problem, given
appropriate programming resource, computing power and computation
time. In other words if the prior distribution and likelihood are
precisely known, then the posterior distribution calculated by
Bayesian inference provides the best possible outcome prediction, as
measured by the highest possible value of the apparent Shannon
information measure, which is then equal to the true mutual
information between input data and the variable being predicted.

Unfortunately in many real life problems neither prior (in the sense
here of the distribution of recurrence time over all patients) nor
likelihood is precisely known. While both can be chosen to reflect the
beliefs of the investigator, there is no guarantee that the
investigator’s beliefs will match reality. Bayesian model choice
provides optimal choice between a number of such mathematical
formulations of prior and likelihood given a dataset, but cannot
itself provide the formulations from which to choose. Under these
circumstances, which match the present case, comparison of algorithms
approximating the true optimal algorithms can be compared using
estimates of the apparent Shannon information content of the
predictions on unseen data\cite{ASI}. Such comparison will place as
joint worst all algorithms which predict to be impossible an outcome
which actually happens.

In this instance, the Cox model is exactly such an algorithm, as it
predicts that relapses can only occur at times at which relapses
occurred in the training set. In order to give the Cox model some
leeway in interpretation, its predictions of relapse time were
deliberately blurred in order to improve its performance under this
very exacting performance measure -- the blurred version turns out to
provide almost exactly zero apparent information about time of relapse
in the 1/2 scenario, a distinct improvement on the $-\infty$ returned
by the unblurred Cox model.

An alternative way of removing the overconfidence of the Cox model is
to remove the assumption that the hazard rate for any patient varies
with time. This results in the constant-hazard Cox model we have
investigated here. As can be seen from its predictions, however, this
model always predicts that the highest hazard of relapse is
immediately after surgery, and that it declines steadily over time --
which is clearly contrary to real-life experience.

The Bayesian model presented here is as far as we know the first such
model for prostate cancer to consider the variation of relapse hazard
with biomarker concentrations to be smooth, as is intuitively
believable. It is also the first model to be shown to provide more
apparent Shannon information than the Cox model and indeed the first
to be shown to provide positive apparent information relative to such
a prior. In all but possibly one of the scenarios
considered it provides more than the constant-hazard variant of the
Cox model also, and in the remaining scenario it does so with
probability 0.79 .

Nonetheless, the total amount of apparent information provided by the
Bayesian model amounts to only 0.1 nepers; in other words, the
uncertainty on time of relapse is reduced on average by a factor of
only about 11\%. Nonetheless the nature of the apparent information
measure is that if one bet successively the entirety of a starting
fund on the predictions holding true according to the scheme detailed
in \cite{ASI} against somebody who knew only the prior, one would make
money at 11\% per patient in the long term without diversification. It
also makes best use of the information that the chosen biomarkers do
provide and gives a platform for improving the biomarker set to
improve the prediction quality.

It is possible that further combinations of input data, prior and
likelihood be presented by ourselves or others in the future, and we
would recommend that the performance of such biomarker and model
combinations is judged against their ability to provide apparent
Shannon information on probability and time of relapse, the measure
which most sensitively distinguishes the performance of prediction
methods at the top end of the quality range. It is likely that given
biomarkers yet to be discovered, the information content of such
predictions will be able to be increased further.

\clearpage 
\appendix

\section{Tables}
\label{tables}

\begin{table}[ht]
\begin{center}
\begin{tabular}{rrrrrrrr}
&&&&\multicolumn{4}{c}{\textbf{Information content (nepers)}}\\
\textbf{Type} & \textbf{Scenario} & \textbf{Method} & \textbf{Concord} & \textbf{0.025q} & \textbf{mean} & \textbf{median} & \textbf{0.975q}\\
time & 1/2 & Bayes & n/a & 0.021 & 0.109 & 0.109 & 0.192 \\
time & 1/2 & Cox & n/a & -0.131 & -0.001 & -0.000 & 0.120 \\
time & 1/2 & CoxCH & n/a & -0.073 & 0.046 & 0.047 & 0.160 \\
order & 1/2 & Bayes & 0.824 & 0.278 & 0.288 & 0.288 & 0.299 \\
order & 1/2 & Cox & 0.795 & 0.165 & 0.181 & 0.181 & 0.196 \\
order & 1/2 & CoxCH & 0.753 & 0.105 & 0.128 & 0.128 & 0.151 \\
time & 10/e & Bayes & n/a & 0.043 & 0.075 & 0.075 & 0.108 \\
time & 10/e & Cox & n/a & -0.131 & -0.022 & -0.015 & 0.047 \\
time & 10/e & CoxCH & n/a & -0.109 & -0.000 & 0.005 & 0.076 \\
order & 10/e & Bayes & 0.839 & 0.302 & 0.308 & 0.308 & 0.313 \\
order & 10/e & Cox & 0.838 & 0.219 & 0.225 & 0.226 & 0.231 \\
order & 10/e & CoxCH & 0.839 & 0.294 & 0.301 & 0.301 & 0.308 \\
time & 8/8 & Bayes & n/a & 0.043 & 0.107 & 0.106 & 0.172 \\
time & 8/8 & Cox & n/a & 0.004 & 0.061 & 0.061 & 0.121 \\
time & 8/8 & CoxCH & n/a & 0.090 & 0.138 & 0.138 & 0.186 \\
order & 8/8 & Bayes & 0.871 & 0.347 & 0.353 & 0.353 & 0.359 \\
order & 8/8 & Cox & 0.839 & 0.213 & 0.219 & 0.219 & 0.224 \\
order & 8/8 & CoxCH & 0.856 & 0.299 & 0.307 & 0.307 & 0.314
\end{tabular}
\caption{Summary of absolute results of each method. The Type column
  indicates whether the prediction was being made on time of relapse
  or pairwise order of relapse. The Scenario column indicates which of
  the scenarios of section \ref{splitting} is being used. The method
  column gives the prediction method used: Bayes, Cox Proportional
  Hazard (leniently interpreted) or Cox Constant Hazard, as described
  in sections \ref{predqual} and \ref{relapsed}. The column headed
  Concord shows the raw concordance rate for predictions of pairwise
  order of relapse. The remaining four columns indicate the apparent
  Shannon information content of the method and its posterior
  quantiles and mean. Note that the $10/e$ scenario involves multiple
  tests on many patients, which have been treated as independent for
  the purposes of analysis, and uses ten different trained
  algorithms. While the 8/8 scenario uses each patient as an unseen
  test patient exactly once, it uses eight different trained
  algorithms.
\label{table1}
}
\end{center}
\end{table}
\clearpage
\begin{table}[ht]
\tiny
\begin{center}
\begin{tabular}{rrrr|r|rrrr|rrrr}
&&&&&\multicolumn{4}{c|}{$A-B$}&\multicolumn{4}{c}{$A/\max(B,0)$}\\
\textbf{Method} $A$ & \textbf{Method} $B$ & \textbf{Scenario} &
\textbf{Type} & \textbf{$P(A>B)$} & \textbf{0.025q} & \textbf{mean} &
\textbf{median} & \textbf{0.975q} & \textbf{0.025q} & \textbf{mean} &
\textbf{median} & \textbf{0.975q}\\
Bayes & Cox & 1/2 & time & 0.927 & -0.039 & 0.110 & 0.110 & 0.265 & 0.526 & Inf & 644.162 & Inf\\
Bayes & CoxCH & 1/2 & time & 0.808 & -0.080 & 0.063 & 0.062 & 0.207 & 0.320 & Inf & 2.293 & Inf\\
CoxCH & Cox & 1/2 & time & 0.710 & -0.123 & 0.047 & 0.047 & 0.221 & -Inf & Inf & 3.278 & Inf\\
Bayes & Cox & 10/e & time & 0.994 & 0.018 & 0.096 & 0.091 & 0.209 & 1.428 & Inf & Inf & Inf\\
Bayes & CoxCH & 10/e & time & 0.956 & -0.009 & 0.075 & 0.071 & 0.186 & 0.873 & Inf & 14.923 & Inf\\
CoxCH & Cox & 10/e & time & 0.634 & -0.109 & 0.021 & 0.021 & 0.154 & -Inf & Inf & 0.575 & Inf\\
Bayes & Cox & 8/8 & time & 0.850 & -0.043 & 0.045 & 0.045 & 0.132 & 0.563 & Inf & 1.749 & 28.213\\
Bayes & CoxCH & 8/8 & time & 0.219 & -0.111 & -0.031 & -0.031 & 0.049 & 0.300 & 0.745 & 0.773 & 1.465\\
CoxCH & Cox & 8/8 & time & 0.973 & -0.001 & 0.076 & 0.077 & 0.152 & 0.987 & Inf & 2.266 & 35.758\\
Bayes & Cox & 1/2 & order & 0.999 & 0.089 & 0.108 & 0.107 & 0.126 & 1.461 & 1.596 & 1.594 & 1.758\\
Bayes & CoxCH & 1/2 & order & 0.999 & 0.136 & 0.160 & 0.160 & 0.186 & 1.910 & 2.262 & 2.255 & 2.761\\
CoxCH & Cox & 1/2 & order & 0.001 & -0.080 & -0.053 & -0.053 & -0.025 & 0.571 & 0.705 & 0.707 & 0.852\\
Bayes & Cox & 10/e & order & 0.999 & 0.075 & 0.083 & 0.083 & 0.091 & 1.324 & 1.367 & 1.366 & 1.414\\
Bayes & CoxCH & 10/e & order & 0.945 & -0.002 & 0.007 & 0.007 & 0.016 & 0.995 & 1.024 & 1.024 & 1.055\\
CoxCH & Cox & 10/e & order & 0.999 & 0.066 & 0.075 & 0.075 & 0.085 & 1.288 & 1.335 & 1.334 & 1.385\\
Bayes & Cox & 8/8 & order & 0.999 & 0.126 & 0.134 & 0.134 & 0.142 & 1.568 & 1.614 & 1.614 & 1.662\\
Bayes & CoxCH & 8/8 & order & 0.999 & 0.037 & 0.046 & 0.046 & 0.056 & 1.118 & 1.151 & 1.151 & 1.186\\
CoxCH & Cox & 8/8 & order & 0.999 & 0.079 & 0.088 & 0.088 & 0.097 & 1.356 & 1.402 & 1.402 & 1.450\\
\end{tabular}
\end{center}
\caption{Comparisons between the different methods. Comparisons in
  terms of delivered apparent Shannon information have been considered
  both as differences and as ratios; in the latter case Inf denotes
  infinite ratio because the inferior method produced zero or negative
  information content.
\label{table2}
}
\end{table}

\section{Harrell's bootstrapping procedure gives wrong results}
\label{bootstrapping}

In this appendix we describe Harrell's bootstrapping procedure for
assessing the quality of a prediction, and give a hypothetical example
illustrating how it gives wrong answers.

\subsection{Harrell's bootstrapping procedure}

Harrell’s Bootstrapping procedure(\cite{Harrell}, page 372), applied
to a general ``measure of correctness'' $\alpha$ of a general
classification algorithm, proceeds as follows:

\begin{enumerate}

\item Collect a dataset $D$; train algorithm on all of $D$; measure
  resulting performance $\alpha$ on all of $D$; call the resulting
  value $A$.

\item \label{draw} Draw a subset $S$ of $D$ of the same size as $D$, with replacement. ($S$ then
usually contains multiple copies of many elements of $D$, and as the number of
elements in $D$ approaches infinity, the fraction of elements of $D$ represented at
least once in $S$ approaches $1-e^{-1}\approx 0.63$.)

\item Train on $S$; measure performance $\alpha$ on $S$; call the
  resulting value $B$.

\item Train on $S$; measure performance $\alpha$ on $D$; call the
  resulting value $C$.

\item \label{opt} Define $O = B - C$ , the ``optimism'' on this subset $S$.

\item \label{iter} Repeat steps \ref{draw} to \ref{opt} a number of times and take
  the average value of $O$.

\item \label{report} Report the measured value of $\alpha$ to be $A - O$.

\end{enumerate}

The intuitive idea here is that measuring performance on the training
set gives overoptimistic results by some value $O$, the
``optimism''. Steps \ref{draw} to \ref{iter} aim to measure $O$, and
step \ref{report} adjusts the value measured on the training set by
subtracting $O$ from it.

In an ideal world we would now list the conditions under which the
above procedure can be shown to give results that are
reliable. However, we are daunted by the length and difficulty of this
task and instead refer the reader to \cite{Austin,Davison}, and give a
very simple example that this procedure gets totally wrong.

\subsection{Example where Harrell's bootstrapping method gives wrong
  results}

This hypothetical example illustrates how Harrell's bootstrapping can
give wildly wrong results.

Suppose we train an algorithm on photos of people, telling it which
are male and which female, but that the algorithm simply memorises
each photo and the corresponding correct answer, then when shown a
photo to classify reports the correct answer if it has seen the photo
before, and otherwise says male or female at random.

Suppose moreover that the population of the world is
infinite\footnote{Uncountably infinite, if you are a mathematician.},
and that a randomly chosen member of the population is male with
probability 0.5 and otherwise female.

Let us now draw a finite training set $D$ of photos at random, so that
approximately half of them will be male, and ``train'' the
``algorithm'' on $D$. The resulting algorithm will clearly be useless
in real life, because the probability that a randomly chosen member of
the infinite population is in the finite set $D$ is zero, so the
probability the algorithm will correctly classify a photo of that
person is $0.5$.

Now after step \ref{draw} of Harrell's procedure, approximately 63\%
of the photos in $D$ will also be in $S$. Therefore we will get the
results $$A=1.0,\ B=1.0,\ C\approx 0.63 +
\frac{1-0.63}{2}\approx 0.82,\ O=B-C\approx0.18,\ A-O\approx0.82$$ so that we end
up reporting that we can get 82\% of photos correctly classified.

However, on encountering a genuinely new photo that the algorithm
hasn't seen before, the probability that it was in $D$ is zero, so the
probability of correctly classifying it is 0.5, so that the true
performance on unseen data is actually on 50\% correct. Thus in this
example Harrell's bootstrapping procedure is seriously misleading.

However, exactly the same result applies to any situation where

\begin{enumerate}

\item The input data tell us only about the classification of exactly
  that point, and nothing about any other point; and

\item There are so many possible data points that we will never see
  the same one twice; and

\item The classification algorithm is flexible enough to effectively memorise the
training set.

\end{enumerate}

\bibliography{ms}
\bibliographystyle{ieeetr}

\end{document}